\documentclass{svproc}
\usepackage{url}
\usepackage{graphicx}%

\begin{document}
\mainmatter              % start of a contribution
\title{Parton Hadron Quantum Molecular Dynamics (PHQMD) --
a Novel Microscopic N-Body Transport Approach for
Heavy-Ion Dynamics and Hypernuclei Production}
\titlerunning{PHQMD}  % abbreviated title (for running head)

\author{E. Bratkovskaya\inst{1,2} \and J. Aichelin\inst{3,4} \and A. Le F\`evre\inst{1} 
\and V. Kireyeu\inst{5} \and V. Kolesnikov\inst{5}  \and Y.~Leifels\inst{1} \and and V. Voronyuk\inst{5}}

\institute{ 
 GSI Helmholtzzentrum f\"ur Schwerionenforschung GmbH,
  Planckstr. 1, 64291 Darmstadt, Germany \\
\email{E.Bratkovskaya@gsi.de},\\  
\and  
 Institut f\"ur Theoretische Physik, Johann Wolfgang Goethe-Universit\"at,
Max-von-Laue-Str. 1, 60438 Frankfurt am Main, Germany \\
\and
SUBATECH, Universit\'e de Nantes, IMT Atlantique, IN2P3/CNRS
4 rue Alfred Kastler, 44307 Nantes cedex 3, France \\
\and
 Frankfurt Institute for Advanced Studies, Ruth Moufang Str. 1, 60438 Frankfurt, Germany \\
\and
 Joint Institute for Nuclear Research, Joliot-Curie 6, 141980 Dubna, Moscow region, Russia}

\maketitle              % typeset the title of the contribution

\begin{abstract}
We present the novel microscopic n-body dynamical transport approach PHQMD 
(Parton-Hadron-Quantum-Molecular-Dynamics) for the description of particle production 
and cluster formation in heavy-ion reactions at relativistic energies.
The PHQMD extends the established PHSD (Parton-Hadron-String-Dynamics) transport approach 
by replacing the mean field by density dependent two body interactions in a similar way 
as in the Quantum Molecular Dynamics (QMD) models.
This allows for the calculation of the time evolution of the n-body Wigner density and therefore
for a dynamical description of clusters and hypernuclei formation. The clusters 
are identified with
the  MST (Minimum Spanning Tree) or the SACA (‘Simulated Annealing Cluster Algorithm’) algorithm which - by regrouping the nucleons
in single nucleons and noninteracting clusters - generates the most bound configuration of nucleons and clusters.
Collisions among particles in PHQMD are treated in the same way as in PHSD. 
In Ref. \cite{Aichelin:2019tnk} we presented the first results from the PHQMD for general 'bulk' observables 
such as rapidity distributions and transverse mass spectra for hadrons 
as well as for clusters production, including hypernuclei, 
at SIS and FAIR/NICA/BES RHIC energies. The selected results on clusters and hypernuclei production from Ref. \cite{Aichelin:2019tnk} are discussed in this contribution.

\end{abstract}
\section{Introduction}
The study of cluster and hypernucleus production, which reflects the phase space density 
during the expansion phase,  is of particular interest from experimental as well as from  theoretical side.
The production mechanisms of hypernuclei may shed light on the theoretical understanding of the  dynamical evolution of heavy-ion reactions which cannot be addressed by other probes.
In particular, 
the formation of heavy projectile/target like hypernuclei elucidates the physics at the transition region
between spectator and participant matter. 
Since hyperons are produced in the overlap region, multiplicity as well as rapidity distributions of hypernuclei formed in the target/projectile region depend crucially on the interactions of the hyperons with the hadronic matter, e.g. cross sections and potentials.
On the other hand, midrapidity hypernuclei test the phase space distribution of baryons in the expanding participant matter,
especially whether the phase space distributions of strange and non-strange baryons are similar  and whether they are in thermal equilibrium.

The description of cluster and hypernuclei formation is a challenging theoretical task
which requires I) the  microscopic dynamical description of the time evolution of heavy-ion collisions; II) the modelling of the mechanisms for the clusters formation.

Cluster formation has often been described either by a coalescence model
\cite{Zhu:2015voa,Feckova:2016kjx} or statistical methods  
\cite{Botvina} assuming that during the heavy-ion reaction at least a subsystem achieves thermal equilibration. 
Both of these models have serious drawbacks. The most essential is that they are not able 
to address the question of  how the clusters are formed and what we can learn from the 
cluster formation about the reaction dynamics. 

\section{PHQMD: basic ideas }

In order to overcome the limitations of existing models for the clusters formation,
we advance the novel Parton-Hadron-Quantum-Molecular 
Dynamics (PHQMD) \cite{Aichelin:2019tnk} approach which is based on the collision integrals of the Parton-Hadron-String
Dynamics approach \cite{Cassing:2009vt,PHSDrev} 
and density dependent 2-body potential interactions of QMD type models \cite{Aichelin:1991xy,Marty:2012vs}. 
These clusters can be identified by two methods:
either by the minimum spanning tree 
(MST) procedure \cite{Aichelin:1991xy} or by a cluster finding algorithm based on 
the simulated annealing technique, 
the Simulated Annealing Clusterization Algorithm (SACA) \cite{Puri:1996qv,Puri:1998te}. Presently 
an extended version -- the  ‘Fragment Recognition In General Application’ (FRIGA) 
\cite{Fevre:2019lll} is under development
which includes symmetry and pairing energy as well as hyperon-nucleon interactions.
The MST algorithm is based on spatial correlations and it is effective in finding 
the clusters at the end of the reaction. 
In order to identify the cluster formation  already at early times of the reaction, 
when the collisions between the nucleons are still on-going  and the nuclear density is high, 
the SACA approach is used. It is based on the idea of 
Dorso and Randrup \cite{Dorso:1992ch} that the most bound configuration of nuclei and nucleons 
evolves in time towards the final cluster distribution.

\section{Results }
First results from combined  PHSD/SACA approach have been reported in \cite{Fevre:2015fua}.
Recently  we presented the first results from the PHQMD approach on 'bulk' dynamics, covering the energy
range from SIS to RHIC, as well as the results on dynamical cluster formation, including hypernuclei,
based on the MST and  SACA models \cite{Aichelin:2019tnk}.
In this contribution we show the selected results from Ref.  \cite{Aichelin:2019tnk}.

\begin{figure}[htp]%
\centering
\includegraphics[scale=.36]{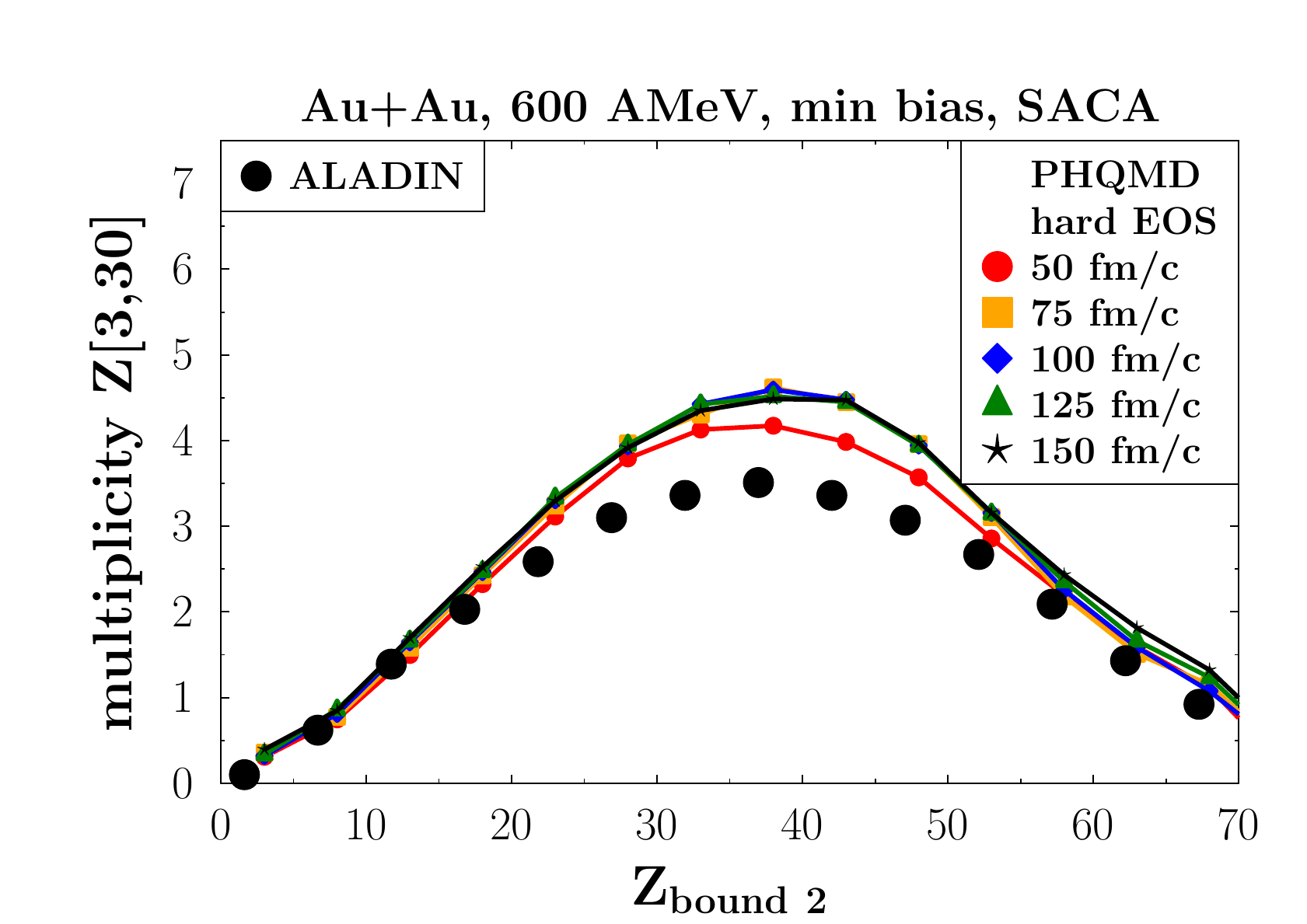}\includegraphics[scale=.36]{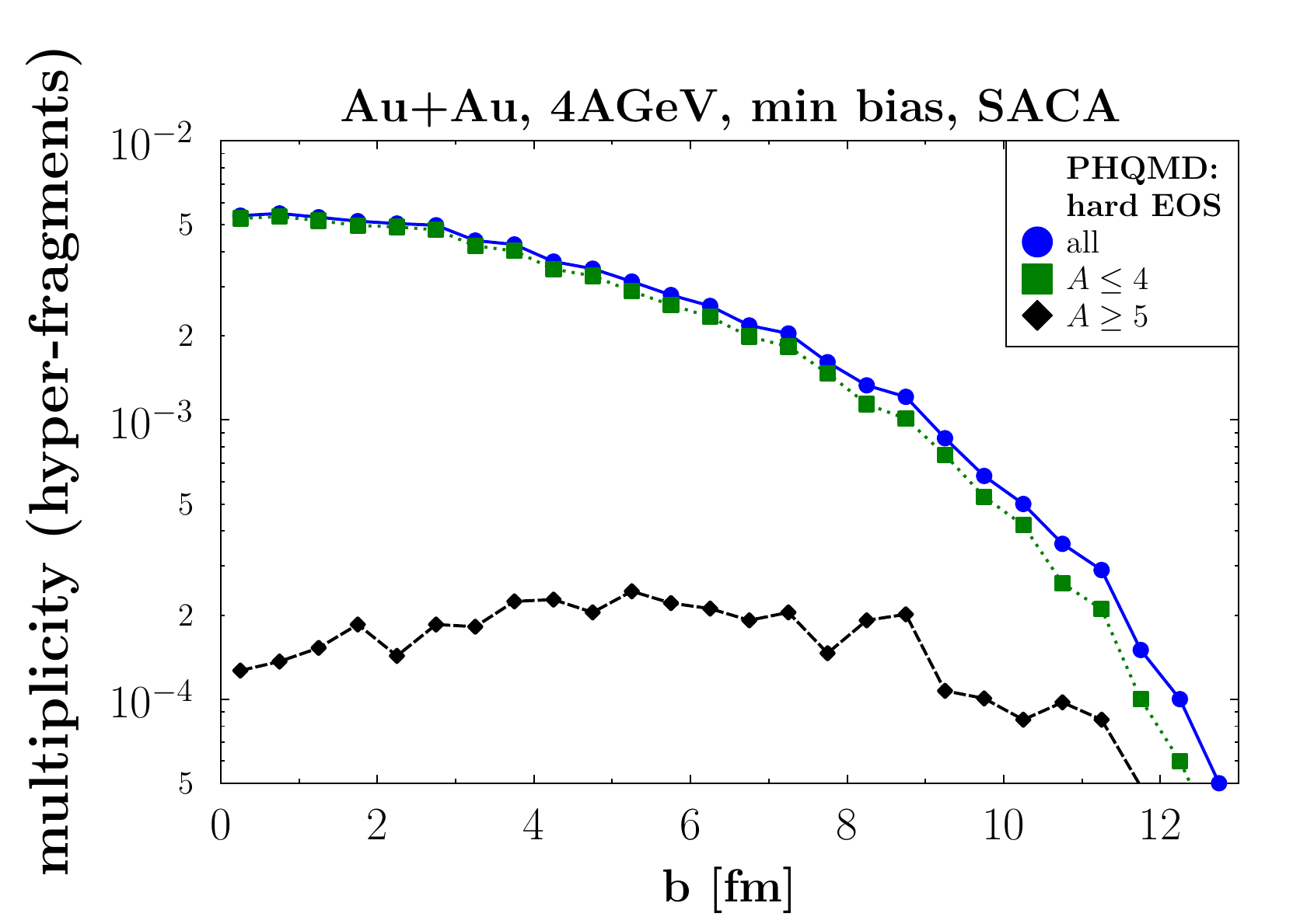}
\caption{\label{rafz} (left) 'Rise and fall' of the multiplicity of clusters with $Z \in [3,30]$ as a function of the total bound charge  $Z_\mathrm{bound\ 2}$.
Both quantities are measured for forward emitted clusters. The experimental data of the ALADIN Collaborations  are from Ref. \cite{LeFevre:2017ygd,Fevre:2019lll}. 
The plot shows the PHQMD results with hard EoS using cluster identification 
by SACA for 600 $A$GeV at different times -- 
50 (red line), 75(orange line), 100 (blue line),  125 (green line), 150 (black line) fm/c.}
\caption{\label{b3} (right)  The multiplicity of hyperclusters as a function of 
the impact parameter for Au+Au collisions at 4 $A$GeV
calculated with the PHQMD using the SACA cluster recognition algorithm.
The blue dots show the multiplicity of all hyper-nuclei, 
while the green squares and black rhombus stand for $A \le 4$ and $A\ge 5$, respectively.} 
\end{figure}

In Fig. \ref{rafz} we display our results for Au+Au at 600 $A$MeV calculated with a hard EoS 
in comparison with minimum bias ALADIN data  \cite{LeFevre:2017ygd}. The clusters identified by SACA are  stable
for times larger than 50 fm/c as shown in Fig. \ref{rafz} .
One can see clearly that PHQMD with a hard EoS reproduces quite nicely the experimentally observed 'rise and fall'.  
The rise and fall of the intermediate mass cluster multiplicity depends strongly on the nuclear equation-of-state. As shown in  \cite{Aichelin:2019tnk},
the rise and fall for a soft EoS overpredicts the data at large  $Z_\mathrm{bound\ 2}$.
There in semi-peripheral and peripheral collisions, where $Z_{bound\ 2}$ is large, the spectator matter is much less stable and fragments into a much larger number of intermediate mass clusters  as compared to a hard EoS (Fig. \ref{rafz}). Thus, the fragment pattern in semi-peripheral reactions can serve as an additional observable to determine the hadronic EoS experimentally.

A special interest is related to the production of hypernuclei in heavy-ion collisions.
In Fig. \ref{b3} we show the multiplicity of light and heavy hypercluster as a function of the 
impact parameter for Au+Au collisions at 4 $A$GeV. 
As seen from this figure, the yield of light hyper-clusters decreases  
with the impact parameter, mainly because the overlap region between projectile and target gets smaller and hence less hyperons are produced. In central collisions,  mainly small hypernuclei ($A \le 4$) are formed while mid-central collisions  
are better suited for a study of heavier hypernuclei ($A \ge 5$). 
Hypernuclei with $A \ge 5$ are dominantly produced by hyperons which enter the spectator matter and get caught there. Therefore, for heavy hyper-nuclei production there is a competition between the hyperon production which decreases with impact parameter and the spectator matter whose size increases with impact parameter.  

Summarizing, we present the PHQMD transport approach which can be used for the dynamical cluster identification including the hypernuclei production from low to ultrarelativistic energies.

\section*{Acknowledgements}
The authors acknowledge inspiring discussions with C. Blume, W. Cassing, C. Hartnack,  
P. Moreau, T. Song, Io. Vassiliev and M. Winn. 
Furthermore, we acknowledge support by the Deutsche Forschungsgemeinschaft 
(DFG, German Research Foundation),  by the Russian Science Foundation grant 19-42-04101
and   by the GSI-IN2P3 agreement under contract number 13-70.
Also we thank the COST Action THOR, CA15213.
The computational resources have been provided by the LOEWE-Center for Scientific Computing
and the "Green Cube" at GSI.


\end{document}